\documentclass[twocolumn,prl,longbibliography]{revtex4-1}

\usepackage{graphicx}
\usepackage{amssymb}
\usepackage{amsmath}
\usepackage{mathtools}
\usepackage{makecell}

\usepackage{multirow}
\usepackage[colorlinks=true,
            linkcolor=red,
            citecolor=blue,
            urlcolor=blue]{hyperref}

\begin{document}

\title{Universal spin dynamics in infinite-temperature one-dimensional quantum magnets}

\author{Maxime Dupont}
\author{Joel E. Moore}
\affiliation{Department of Physics, University of California, Berkeley, California 94720, USA}
\affiliation{Materials Sciences Division, Lawrence Berkeley National Laboratory, Berkeley, California 94720, USA}

\begin{abstract}
    We address the nature of spin dynamics in various integrable and nonintegrable, isotropic and anisotropic quantum spin-$S$ chains, beyond the paradigmatic $S=1/2$ Heisenberg model. In particular, we investigate the algebraic long-time decay $\propto t^{-1/z}$ of the spin-spin correlation function at infinite temperature, using state-of-the-art simulations based on tensor network methods. We identify three universal regimes for the spin transport, independent of the exact microscopic model: (i) superdiffusive with $z=3/2$, as in the Kardar-Parisi-Zhang universality class, when the model is integrable with extra symmetries such as spin isotropy that drive the Drude weight to zero, (ii) ballistic with $z=1$ when the model is integrable with a finite Drude weight, and (iii) diffusive with $z=2$ with easy-axis anisotropy or without integrability, at variance with previous observations.
\end{abstract}

\maketitle

\textit{Introduction.} Understanding equilibrium and out-of-equilibrium dynamics of interacting quantum systems remains one of the most strenuous problems in modern physics. From a phenomenological perspective, taking into account the few conservation laws of a system such as energy, momentum, and particle number, one can derive classical hydrodynamic equations to describe a coarse-grained thermodynamic version of the microscopic model~\cite{kadanoff1963,landau1987}. Yet, some systems possess an extensive set of conservation laws, strongly constraining their dynamics and endowing them with exotic thermalization and transport properties~\cite{prosen2011,caux2013,wouters2014,ilievski2015,essler2016,ilievski2016}. They are known as \textit{integrable} systems and are ubiquitous in the low-dimensional quantum world, with experimentally relevant examples from magnets to Bose gases~\cite{lieb1963,giamarchi2004,kinoshita2006,hild2014,langen2015,tang2018}.

Two simple paradigms of how a conserved quantity spreads are exemplified by ordinary thermalizing systems with diffusion on the one hand, and free-particle systems (a simple kind of integrable system) with ballistic transport on the other. After many years and much analytical and numerical progress~\cite{zotos1997,sirker2006,sirker2011,prosen2011,karrasch2013,ilievski2015,bertini2016,bulchandani2018,denardis2018,gopalakrishnan2018,denardis2019scipost,agrawal2019}, the existence of both regimes in the spin-half XXZ model, which is a version of the Heisenberg model with uniaxial anisotropy in the interaction, has been understood in detail, with quantitative explanations of the Drude weight that governs the amount of ballistic transport. Numerical studies on this model provide a stringent test of the generalized hydrodynamical approach to time evolution of densities in ballistic regimes of integrable models~\cite{castroalvaredo2016,bertini2016,bulchandani2018}.

Unexpectedly, a numerical study observed a third behavior at the isotropic (Heisenberg) point of this model~\cite{znidaric2011,ljubotina2019}: spin dynamics at infinite temperature were characterized by superdiffusion with the same dynamical critical exponent $z=3/2$, defined below, that appears in the \textit{classical}, \textit{stochastic} Kardar-Parisi-Zhang (KPZ) universality class~\cite{kardar1986}. This led to additional studies that explained how the diffusion constant must become infinite at the Heisenberg point~\cite{gopalakrishnan2019} and showed agreement with the full KPZ scaling function~\cite{ljubotina2019,gopalakrishnan2019b,krajnik2019,spohn2019,weiner2020}. Note that this emergence of superdiffusion and KPZ universality from quantum models is different from the superdiffusion with $z=1$ that emerges in systems with momentum conservation~\cite{narayan2002,gao2017} or the variable dynamical critical exponent at low temperatures in Luttinger liquids~\cite{bulchandani2019}.  It also does not seem to follow from the useful mapping between a classical exclusion process in the KPZ universality class and \textit{statics} of the spin-half XXZ model (for a review, see e.g., Ref.~\onlinecite{quastel2015}).

The main point of this Rapid Communication is to study infinite-temperature dynamics in a variety of one-dimensional quantum magnets with $S>1/2$, with and without integrability and isotropy, in order to isolate the requirements for KPZ superdiffusion. We find several new examples of higher-spin chains that all have dynamical critical exponent $z=3/2$, despite having variable symmetries and interactions. These can be viewed as interpolating between the $S=1/2$ results and recent studies of a classical integrable spin chain~\cite{das2019}. We find that the occurrence of superdiffusion with $S=1$ is not limited to the isotropic case, but that it does require integrability; more precisely, we find that superdiffusion is not present in the simplest nearest-neighbor models with $S=1$, $3/2$ and $2$, contrary to recent proposals~\cite{denardis2019}, and we explain what we believe to be missing in that theoretical analysis.

\textit{Investigating spin dynamics.} To investigate the spin dynamics in quantum spin-$S$ systems, we focus on the infinite-temperature local spin-spin correlation function,
\begin{equation}
    \mathcal{C}(L,t)=\frac{3}{S(S+1)}\;\Bigl\langle S^z_{L/2}(t)\; S^z_{L/2}(0)\Bigr\rangle,
    \label{eq:autocorrelation_function}
\end{equation}
where $S^z_{L/2}$ is the spin operator component along the quantization axis at position $L/2$ in a system of total length $L$, $\langle\cdot\rangle\equiv\mathrm{tr}(\cdot)/(2S+1)^L$ denotes the infinite temperature thermal average, and $S^z_r(t)=\mathrm{e}^{i\mathcal{H}t}S^z_r\mathrm{e}^{-i\mathcal{H}t}$ is the time-dependent operator in the Heisenberg picture, with $\mathcal{H}$ the Hamiltonian describing the system. The prefactor $3/S(S+1)$ in Eq.~\eqref{eq:autocorrelation_function} ensures that $\mathcal{C}(L,t=0)=1$.

\begin{figure*}[!t]
    \centering
    \includegraphics[width=2\columnwidth]{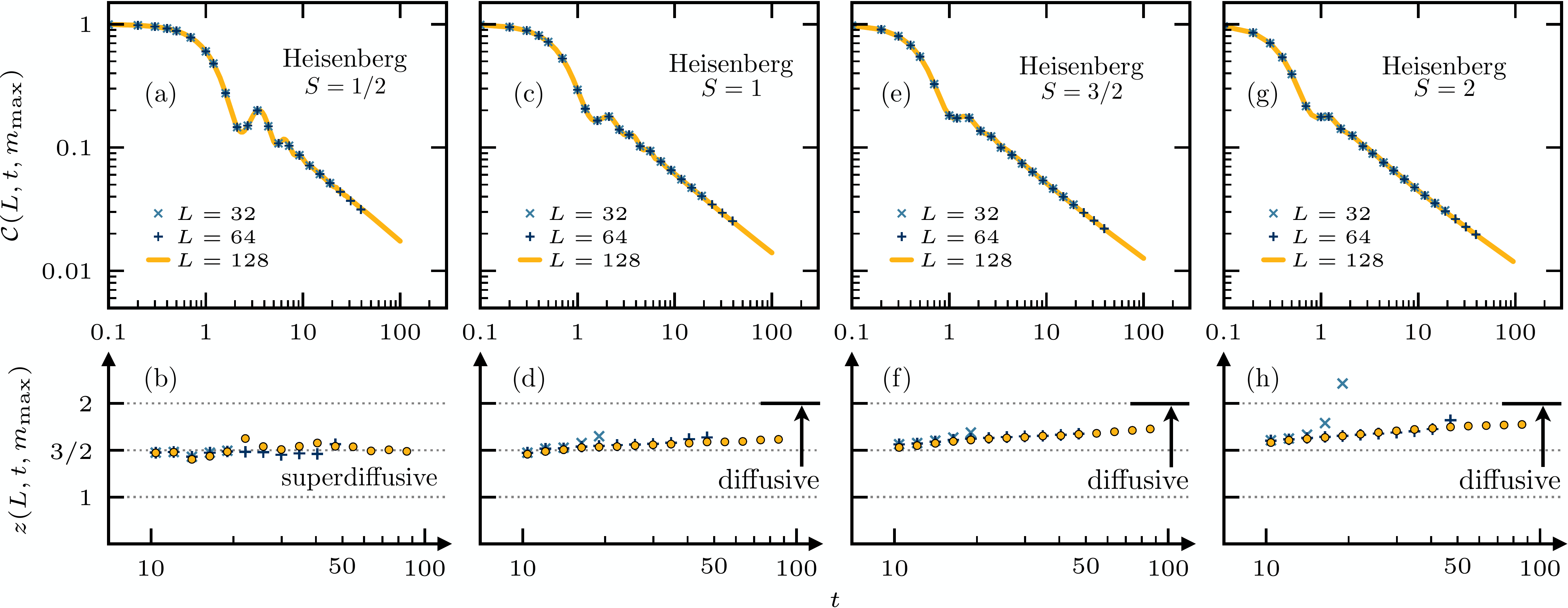}
    \caption{Top panels~(a,\,c,\,e,\,g): Infinite temperature spin-spin correlation function~\eqref{eq:autocorrelation_function} for the isotropic one-dimensional Heisenberg model~\eqref{eq:ham_heisenberg} for spin values $S=1/2$, $1$, $3/2$ and $2$. Bottom panels~(b,\,d,\,f,\,h): Extracted dynamical exponent $z(L,t,m_\mathrm{max})$ by performing curve-fitting inside a sliding window of data points in order to reliably extract the infinite-length and infinite-time value of the power-law decay as in Eq.~\eqref{eq:algebraic_decay}. Only the spin-$1/2$ case is integrable and shows consistent superdiffusive dynamical behavior over time with $z(L,t,m_\mathrm{max})=3/2$. For the larger spin-$S$ models, the dynamical exponent value systematically increases when varying the curve-fitting window toward longer times with $z\to 2$, supporting diffusive dynamics. Additional analyses are available in the Supplemental Material.}
    \label{fig:heisenberg_model}
\end{figure*}

We consider a wide range of integrable and nonintegrable, isotropic and anisotropic quantum spin-$S$ chains described by Hamiltonians of the form $\mathcal{H}=\sum_j\hat{h}_{j,j+1}$ with $\hat{h}_{j,j+1}$ the local Hamiltonian density. All models conserve the total magnetization $S^z_\mathrm{tot}=\sum_j S^z_j$, and some additionally conserve the total spin $\boldsymbol{S}_\mathrm{tot}=\sum_j \boldsymbol{S}_j$, where $\boldsymbol{S}_j=\left(S^x_j,S^y_j,S^z_j\right)$ is the usual spin-$S$ operator at site $j$, making them fully isotropic. Because of $S^z_\mathrm{tot}$ conservation, in the hydrodynamic limit, the spin fluctuations captured by the spin-spin correlation function~\eqref{eq:autocorrelation_function} are expected to decay with a power-law tail at late time for infinitely large systems,
\begin{equation}
    \lim_{t\to\infty}\lim_{L\to\infty}~\mathcal{C}(L,t)\;\sim\; t^{-1/z},
    \label{eq:algebraic_decay}
\end{equation}
with $z$ the dynamical exponent characterizing the nature of the spin dynamics and spin transport in the system: $z=2$ for diffusion, $z=3/2$ for KPZ-type anomalous diffusion or superdiffusion, and $z=1$ for ballistic dynamics.

We compute the spin-spin correlation function~\eqref{eq:autocorrelation_function} numerically using matrix product states (MPS) calculations~\cite{schollwock2011} together with the purification method~\cite{verstraete2004}. The time evolution is performed through the time-evolving block decimation algorithm~\cite{vidal2004} along with a fourth order Trotter decomposition~\cite{hatano2005} of time step $\delta_t=0.1$. The control parameter of the numerical simulations is the bond dimension $m$ of the MPS whose convergence is thoroughly studied in the Supplemental Material~\cite{supplemental}. In the following, we only show data for the largest bond dimension computationally available, $m\equiv m_\mathrm{max}$. In practice one only has access to finite systems $L$ and is limited in the maximum time $t$. Therefore, it is instructive to perform curve-fitting inside a sliding window of data points in order to reliably extract the infinite-length and infinite-time value of the dynamical exponent $z$~\cite{supplemental}.

\textit{The Heisenberg model.}---We first consider the paradigmatic $\mathrm{SU}(2)$-symmetric Heisenberg model,
\begin{equation}
    \hat{h}_{j,j+1}=\boldsymbol{S}_j\cdot\boldsymbol{S}_{j+1},
    \label{eq:ham_heisenberg}
\end{equation}
for $S=1/2$, $1$, $3/2$ and $2$, and which is integrable exclusively in the spin-$1/2$ case~\cite{bethe1931,franchini2017}. The correlation function~\eqref{eq:autocorrelation_function} and the extracted dynamical exponent $z$ are shown in Fig.~\ref{fig:heisenberg_model}. Superdiffusive behavior with $z=3/2$ is unambiguously observed for the $S=1/2$ case, in agreement with previous results~\cite{znidaric2011,znidaric2011b,ilievski2018,ljubotina2019,gopalakrishnan2019,gopalakrishnan2019b}. $z=3/2$ is the same dynamical scaling exponent as of the KPZ universality class~\cite{kardar1986}, and the relation has been confirmed by showing that the infinite temperature spin-spin correlation function obeys KPZ scaling~\cite{ljubotina2019,gopalakrishnan2019b}. For larger spins $S\geq 1$, the dynamical exponent value systematically increases when varying the curve-fitting window toward longer times with $z\to 2$, supporting diffusive dynamics.

\begin{figure*}[t!]
    \centering
    \includegraphics[width=2\columnwidth]{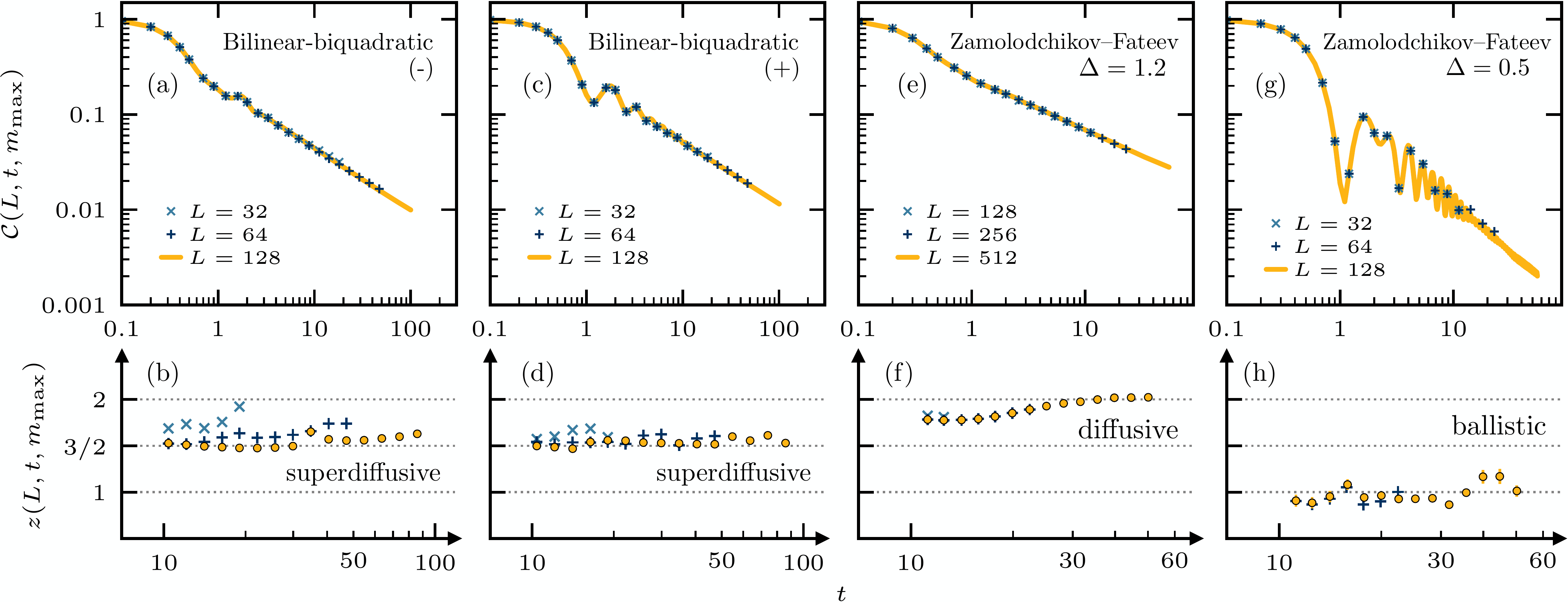}
    \caption{Top panels~(a,\,c,\,e,\,g): Infinite temperature spin-spin correlation function~\eqref{eq:autocorrelation_function} for various one-dimensional spin-$1$ models. First, the isotropic bilinear-biquadratic Heisenberg chain defined in Eq.~\eqref{eq:ham_bibi} for two different signs of the biquadratic term (panels a and c respectively). Then, the Zamolodchikov-Fateev model~\eqref{eq:ham_zf} with easy-plane ($\Delta=0.5$) and easy-axis ($\Delta=1.2$) anisotropy (panels e and g respectively). Bottom panels~(b,\,d,\,f,\,h): Extracted dynamical exponent $z(L,t,m_\mathrm{max})$ by performing curve-fitting inside a sliding window of data points. Superdiffusion is observed at the isotropic $\mathrm{SU}(2)$ and $\mathrm{SU}(3)$ points of the bilinear-biquadratic $S=1$ chain while diffusive and ballistic spin dynamics are respectively obtained for the easy-axis and easy-plane ZF model. Additional analyses are available in the Supplemental Material.}
    \label{fig:spin_one_models}
\end{figure*}

Our result of diffusive dynamics in these nonintegrable cases is perhaps not too surprising, but there is a relatively long crossover before reaching this limit, and the fact that at short time $z\approx 3/2$ can be misleading. For instance, based on calculations on a low-energy effective quantum field theory for the Heisenberg model~\eqref{eq:ham_heisenberg}, namely, the non-linear sigma model~\cite{haldane1983,haldane1983b,affleck1987}, the authors of Ref.~\onlinecite{denardis2019} claim that anomalous spin transport is present in any spin-$S$ Heisenberg chain at low temperature, and persists at high temperature as corroborated by simulations on the exact spin-$1$ microscopic model. However, their simulations do not go to long enough time to observe the increase of $z$ as we do. The superdiffusive dynamics that they obtain is an artifact of the low-energy field theory which is integrable~\cite{zamolodchikov1979,zamolodchikov1992}, while the exact microscopic model is not.  This long-time crossover to diffusion could possibly have been anticipated based on previous studies on integrability breaking in $S=1/2$ quantum spin chains, where the integrability breaking is controlled either by adding a parameter or by going to low temperature~\cite{sirker2011,huang2013}. For example, the charge conductivity is finite with broken integrability but diverges as a powerlaw in inverse temperature or strength of integrability breaking~\cite{huang2013}, because of the same kind of long-time crossover observed here. The result of diffusion in the nonintegrable $S\geq 1$ Heisenberg chain is further evidence that integrability breaking should be regarded as a ``dangerously irrelevant'' perturbation to dynamics at long times~\cite{vasseur2016}: even if the breaking is weak and irrelevant at low energy in the renormalization group sense, it can strongly modify the long-time behavior by inducing thermalization. It is worth noting two other recent works mentioning (super)diffusion in the $S=1$ Heisenberg chain~\cite{capponi2019,richter2019b}, although they could not provide a definitive answer regarding the nature of the spin dynamics.

Even in the classical limit $S\to\infty$, where spin operators in Eq.~\eqref{eq:ham_heisenberg} are replaced by standard unit vectors, identifying whether spin diffusion is normal or anomalous has a long-standing history~\cite{muller1988,gerling1989,muller1988,liu1991,alcantara1992,bohm1993,lovesey1994,lovesey1994b,srivastava1994}. The issue was settled by doing a systematic finite-size analysis in Ref.~\onlinecite{bagchi2013}: As in the quantum cases displayed in Fig.~\ref{fig:heisenberg_model}, $z\to 2$ is only reached asymptotically at relatively long time. This confirms normal diffusive spreading of spin fluctuations, as expected for a nonintegrable model. Interestingly, the spin dynamics of an integrable \textit{classical} spin chain with the same symmetries as the Heisenberg model, known as the Faddeev-Takhtajan model~\cite{faddeev2007,avan2010,prosen2013}, has recently been explored~\cite{das2019}. The authors are able to show that the spin transport is superdiffusive with $z=3/2$, and belongs to the KPZ universality class, just like the quantum spin-$1/2$ Heisenberg chain. In addition to the isotropic point, easy-plane and easy-axis regimes of the model are also investigated and respectively exhibit ballistic and diffusive spin transport; again, just like the quantum $S=1/2$ Heisenberg model. This legitimately raises questions of possible universality regarding the spin dynamics depending on the nature of the anisotropy in the model. To address this, we extend the current study to larger spin-$S$ quantum models.

\textit{Family of S=1 models.} We first turn our attention to various spin-$1$ models, starting with the isotropic bilinear-biquadratic Heisenberg chain,
\begin{align}
    \hat{h}_{j,j+1}=\boldsymbol{S}_j\cdot\boldsymbol{S}_{j+1}\pm \left(\boldsymbol{S}_j\cdot\boldsymbol{S}_{j+1}\right)^2.
    \label{eq:ham_bibi}
\end{align}
The two cases considered, with the $\pm$ sign for the biquadratic term, are both integrable. With the minus sign, the model is known as the $\mathrm{SU}(2)$-invariant Babujian-Takhtajan Hamiltonian~\cite{takhtajan1982,babujian1982,babujian1983}. Its dynamical spin-spin correlation function~\eqref{eq:autocorrelation_function} as well as the long-time decay exponent $z$ are plotted in Fig.~\ref{fig:spin_one_models}\;(a,\,b) and show superdiffusion. Here, anomalous spin dynamics is observed in a quantum magnet besides the spin-$1/2$ Heisenberg chain, and might hint that something universal is responsible for this behavior in integrable systems, such as the rotation symmetry. This is why the Hamiltonian~\eqref{eq:ham_bibi} with a plus sign (known as the Uimin-Lai-Sutherland model~\cite{uimin1970,lai1974,sutherland1975}) is interesting, because it extends the $\mathrm{SU}(2)$ symmetry to $\mathrm{SU}(3)$, and still demonstrates superdiffusive spin dynamics, see Fig.~\ref{fig:spin_one_models}\;(c,\,d). This means that having an integrable $\mathrm{SU}(2)$-symmetric model is not in itself a necessary ingredient to have anomalous diffusion, as pointed out in Ref.~\onlinecite{ilievski2018}. This statement will be extended by looking at an integrable $\mathrm{SO}(5)$-symmetric spin-$2$ chain.

Before that, to investigate the effect of anisotropy, we consider the anisotropic $S=1$ Zamolodchikov-Fateev (ZF) model~\cite{zamolodchikov1980},
\begin{align}
    \hat{h}_{j,j+1}={}&~\quad S^x_jS^x_{j+1} + S^y_jS^y_{j+1} +\left(2\Delta^2-1\right) S^z_jS^z_{j+1}\nonumber\\
    &+~ 2\left[\left(S^x_j\right)^2+ \left(S^y_j\right)^2+\left(2\Delta^2-1\right)\left(S^z_j\right)^2\right]\nonumber\\
    &- \sum_{a,b\in[x,y,z]} f_{ab}(\Delta) S^a_jS^a_{j+1}S^b_jS^b_{j+1},
    \label{eq:ham_zf}
\end{align}
where $f_{ab}=f_{ba}$, $f_{zz}=2\Delta^2-1$, $f_{xz}=f_{yz}=2\Delta-1$ and $f_{ab}=1$ otherwise. This model is analogous to the quantum spin-$1/2$ XXZ chain in the sense that it is parametrized by a continuous anisotropy parameter $\Delta$ and that it is integrable~\cite{sogo1984,kirillov1986,mezincescu1990}. At the isotropic point $\Delta=1$, it coincides with the Babujian-Takhtajan Hamiltonian~\eqref{eq:ham_bibi} previously studied. In the presence of easy-axis anisotropy, i.e., $|\Delta|>1$, we observe diffusive dynamics, as shown in Fig.~\ref{fig:spin_one_models}\;(e,\,f) for $\Delta=1.2$, while for an easy-plane anisotropy $|\Delta|=0.5<1$, dynamics is ballistic. In the latter case, ballistic transport is expected as the Mazur bound~\cite{mazur1969,suzuki1971,zotos1997} computed analytically in Ref.~\onlinecite{piroli2016} establishes a nonvanishing Drude weight for this model.

\begin{figure}[t!]
    \centering
    \includegraphics[width=\columnwidth]{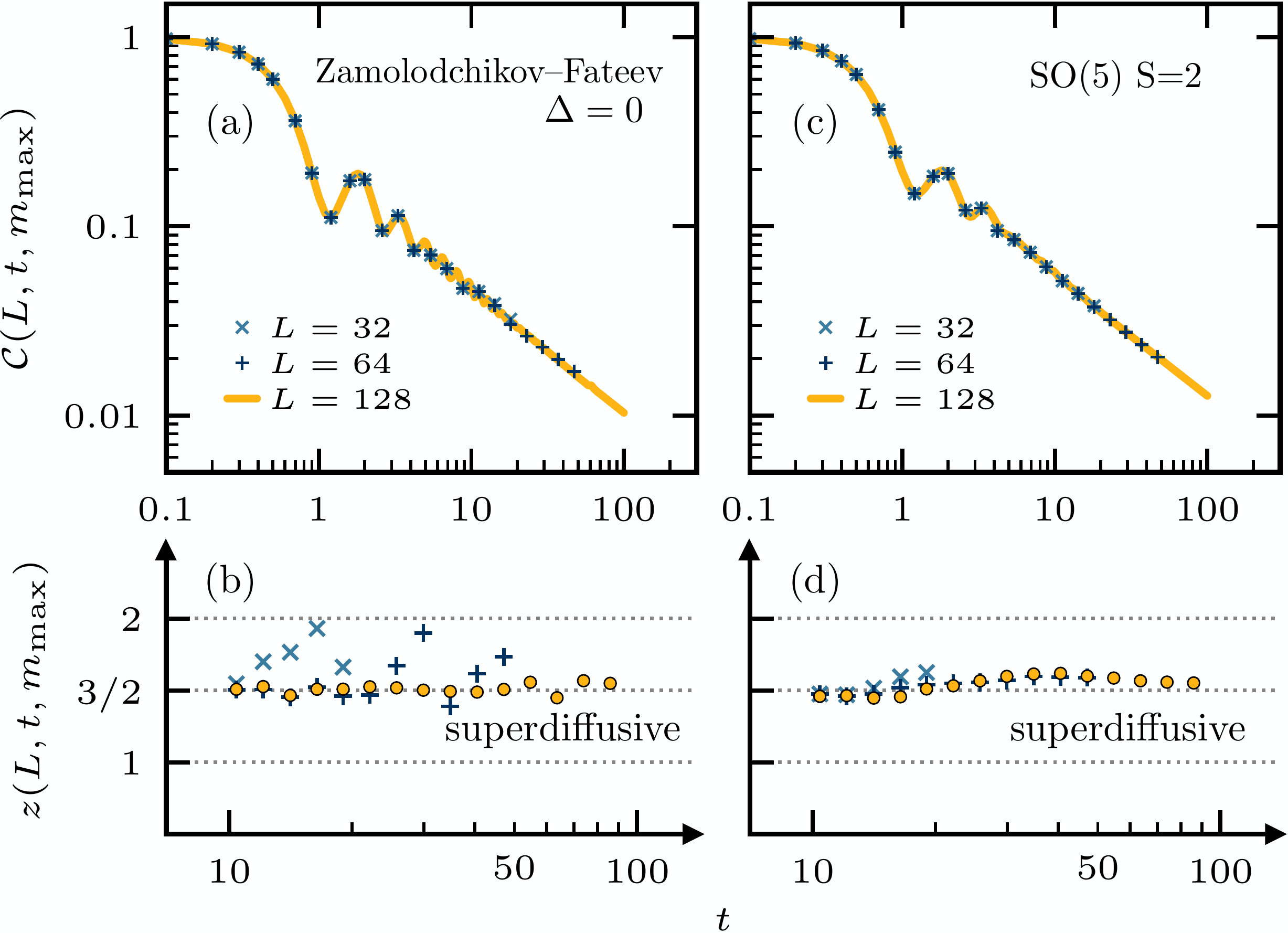}
    \caption{Top panels~(a,\,c): Infinite temperature spin-spin correlation function~\eqref{eq:autocorrelation_function} for the spin-$1$ Zamolodchikov-Fateev model~\eqref{eq:ham_zf} at $\Delta=0$ and the $\mathrm{SO}(5)$-symmetric bilinear-biquadratic $S=2$ model defined in Eq.~\eqref{eq:ham_so5} for $\theta=\arctan(1/9)$. Bottom panels~(b,\,d): Extracted dynamical exponent $z(L,t,m_\mathrm{max})$ by performing curve-fitting inside a sliding window of data points. Superdiffusive spin dynamics is observed in both cases with $z=3/2$. Additional analyses are available in the Supplemental Material.}
    \label{fig:extra_models}
\end{figure}

Overall, the dependence on the spin dynamics (diffusive, ballistic and superdiffusive) on the anisotropy is quite familiar, with identical behavior observed for the spin-$1/2$ quantum Heisenberg chain~\cite{znidaric2011}, the classical Faddeev-Takhtajan model~\cite{das2019}, and now the $S=1$ ZF model. However, an interesting feature at $S=1$ is that the ZF model also shows superdiffusion in the ``easy-plane limit'' $\Delta=0$, see Fig.~\ref{fig:extra_models}\;(a,\,b). The possibility that the $\Delta=0$ point in the ZF model is special was previously pointed out~\cite{piroli2016} on the grounds that it is not forced to have ballistic transport by the conserved quantities that force nonzero Drude weight at other values $0<|\Delta|<1$.

\textit{Integrable SO(5)-symmetric spin-2 chain.} To confirm the universal nature of superdiffusion in integrable isotropic magnets, we study a generalization of the $S=1$ bilinear-biquadratic Heisenberg chain~\eqref{eq:ham_bibi}. It can be written down as a one-parameter family of bilinear-biquadratic Hamiltonians in terms of the $\mathrm{SO}(2n+1)$ generators~\cite{tu2008,tu2008b}. Focusing on the $n=2$ case~\footnote{In fact, we have already studied the spin dynamics of the most interesting points of the $n=1$ case, which can be expressed as a spin-$1$ model. These points correspond to the Babujian–Takhtajan and Uimin-Lai-Sutherland Hamiltonians defined in Eq.~\eqref{eq:ham_bibi}.} and using a spin-$2$ formulation of this model~\cite{tu2008,tu2008b,alet2011}, one gets,
\begin{align}
    \hat{h}_{j,j+1}={}&\quad\cos\theta\Bigl[-1-\frac{5}{6}\boldsymbol{S}_j\cdot\boldsymbol{S}_{j+1} + \frac{1}{9}\left(\boldsymbol{S}_j\cdot\boldsymbol{S}_{j+1}\right)^2\nonumber\\
    &+\frac{1}{18}\left(\boldsymbol{S}_j\cdot\boldsymbol{S}_{j+1}\right)^3\Bigr] + \sin\theta\Bigl[1 - 5\boldsymbol{S}_j\cdot\boldsymbol{S}_{j+1}\nonumber\\
    &-\frac{17}{12}\left(\boldsymbol{S}_j\cdot\boldsymbol{S}_{j+1}\right)^2 + \frac{1}{3}\left(\boldsymbol{S}_j\cdot\boldsymbol{S}_{j+1}\right)^3 \nonumber\\
    &+\frac{1}{12}\left(\boldsymbol{S}_j\cdot\boldsymbol{S}_{j+1}\right)^4\Bigr].
    \label{eq:ham_so5}
\end{align}
It has an integrable point at $\theta=\arctan(1/9)$, as well as other remarkable points whose values can be generalized as a function of $n$ for all symmetry groups~\cite{reshetikhin1983,reshetikhin1985,affleck1991,scalapino1998,alet2011}. We show in Fig.~\ref{fig:extra_models}\;(c,\,d) that, once more, anomalous diffusion is present at an integrable and isotropic point which is neither characterized by $\mathrm{SU}(2)$, nor $\mathrm{SU}(3)$ but $\mathrm{SO}(5)$ in this case.

\textit{Summary and discussions.} Employing extensive numerical simulations based on tensor network methods, we have investigated the algebraic long-time decay of the infinite temperature spin-spin correlation function in various integrable and nonintegrable, isotropic and anisotropic quantum spin-$S$ chains~\cite{supplemental}. Our results unequivocally support universal spin dynamics in infinite-temperature one-dimensional magnets, with three different possible regimes: (i) superdiffusive, as in the KPZ universality class, when the model is integrable with extra symmetries such as spin isotropy that drive the Drude weight to zero, (ii) ballistic when the model is integrable with a finite Drude weight, and (iii) diffusive otherwise.

One potential future direction is to demonstrate that the full KPZ~\cite{kardar1986} scaling function $f_\mathrm{KPZ}$ is indeed present for all models showing anomalous diffusion, i.e., $\langle S^z_r(t)S^z_0(0)\rangle\sim t^{-2/3}\,f_\mathrm{KPZ}[r(\lambda t)^{-2/3}]$ with $\lambda$ some parameter~\cite{prahofer2004,spohn2014}. As it is very costly to compute the dynamical spin-spin correlation function at all distances $r$, it would be numerically preferable to use the workaround developed in Ref.~\onlinecite{ljubotina2019} for $S=1/2$ to address this question. An open puzzling question is what ingredient(s) makes the superdiffusive behavior with $z=3/2$ robust in all isotropic integrable magnets, classical and arbitrary spin-$S$ quantum models alike? It would also be interesting to see if the mechanism of anomalous diffusion proposed in Ref.~\onlinecite{gopalakrishnan2019} for the spin-half Heisenberg chain can be extended to all these superdiffusive examples.

\begin{acknowledgments}
    \textit{Acknowledgments.} M.D. is grateful to S. Capponi, M. Schmitt, and J. Wurtz for interesting discussions at the early stage of this work. We also acknowledge discussions with Z. Lenar{\v c}i{\v c}, V. Bulchandani, S. Gopalakrishnan, C. Karrasch, and J. De Nardis. This work was funded by the U.S. Department of Energy, Office of Science, Office of Basic Energy Sciences, Materials Sciences and Engineering Division under Contract No. DE-AC02-05-CH11231 through the Scientific Discovery through Advanced Computing (SciDAC) program (KC23DAC Topological and Correlated Matter via Tensor Networks and Quantum Monte Carlo). J.E.M. acknowledges support from a Simons Investigatorship. This research used the Lawrencium computational cluster resource provided by the IT Division at the Lawrence Berkeley National Laboratory (supported by the Director, Office of Science, Office of Basic Energy Sciences, of the U.S. Department of Energy under Contract No. DE-AC02-05CH11231). This research also used resources of the National Energy Research Scientific Computing Center (NERSC), a U.S. Department of Energy Office of Science User Facility operated under Contract No. DE-AC02-05CH11231. The code for calculations is based on the ITensor library~\footnote{ITensor library, \href{http://itensor.org}{http://itensor.org}.}.
\end{acknowledgments}

\bibliography{bibliography}

\appendix
\setcounter{figure}{0}
\setcounter{equation}{0}
\newpage\clearpage\onecolumngrid

\begin{center}
    \large\textbf{Supplemental material to ``\textit{Universal spin dynamics\\ in infinite-temperature one-dimensional quantum magnets}''}
\end{center}

\begin{center}
    \begin{minipage}{0.9\textwidth}
    Firstly, we show that the fitting procedure to extract the dynamical exponent $z$ is stable by considering different sizes for the fitting window. Secondly, we discuss the finite-size effects which are visible in the plots of the main text. Then, we show that good convergence of our results versus the bond dimension $m$ of the matrix product states is achieved. Finally, four additional models are studied, supporting our conclusions. Specifically, we look at an integrable $\mathrm{SU}(5)$ spin-$2$ model, the non-integrable XY $S=1$ and $S=3/2$ models and the isotropic dimerized spin-$1$ chain. A summary of the parameters for all the models considered is also available.
    \end{minipage}
\end{center}

\section*{Robustness of the fitting procedure}

\begin{figure*}[!h]
    \centering
    \includegraphics[width=\columnwidth]{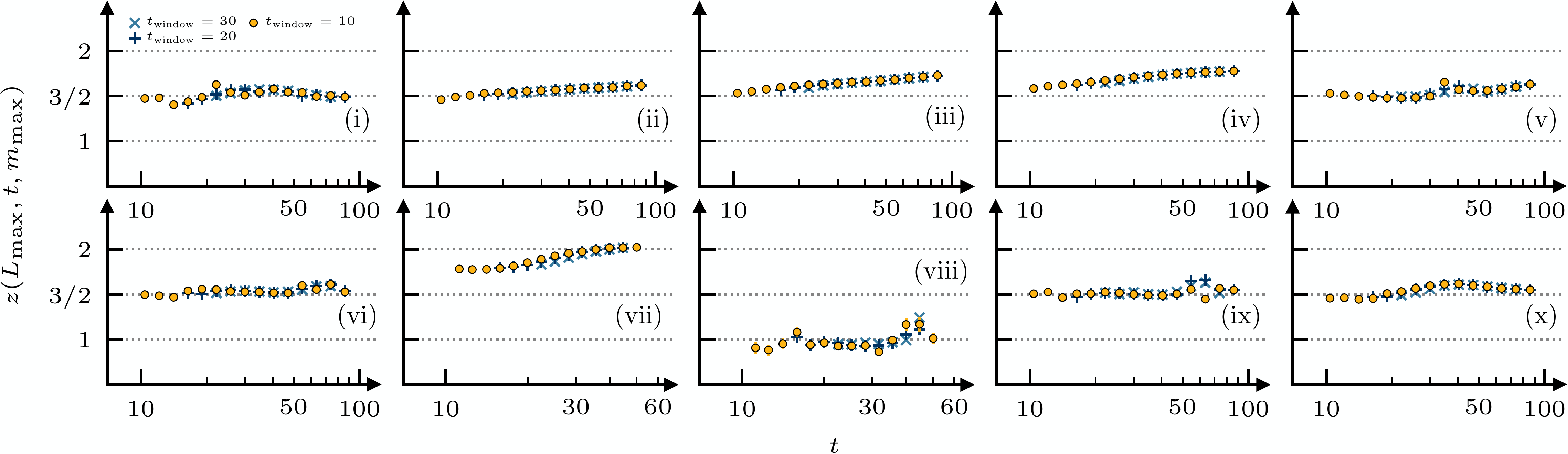}
    \caption{Extracted dynamical exponent $z$ by considering different sizes ($t=10$, $t=20$ and $t=30$) for the fitting window. Each model of the main text is considered at the largest system size $L_\mathrm{max}$ and largest bond dimension $m_\mathrm{max}$ (see Tab.~\ref{tab:sim_parameters}). The panels correspond to: (i) Heisenberg $S=1/2$, (ii) Heisenberg $S=1$, (iii) Heisenberg $S=3/2$, (iv) Heisenberg $S=2$, (v) Babujian-Takhtajan $S=1$, (vi) Uimin-Lai-Sutherland $S=1$, (vii) Zamolodchikov-Fateev $S=1$ at $\Delta=0.5$, (viii) Zamolodchikov-Fateev $S=1$ at $\Delta=1.2$, (ix) Zamolodchikov-Fateev $S=1$ at $\Delta=0.0$, (x) $\mathrm{SO}(5)$ symmetric $S=2$ model.}
    \label{fig:fit_window_size}
\end{figure*}

To evaluate the robustness of the fitting procedure and reliably extract the dynamical exponent $z$, we try different size for the fitting window: $t=10$, $20$ and $30$. Each window contains $10t$ data points because of the Trotter time step $\delta_t=0.1$ considered to perform the time evolution. The largest system size $L_\mathrm{max}$ and largest bond dimension $m_\mathrm{max}$ of each model of the main text is considered in Fig.~\ref{fig:fit_window_size}. We see that the fitting procedure is stable with no deviation for $z$ versus the size of the fitting window. The extracted dynamical exponents $z$ in the main text correspond to a time window of size $t=10$.

\section*{Convergence with the bond dimension}

\begin{figure*}[!t]
    \centering
    \includegraphics[width=\columnwidth]{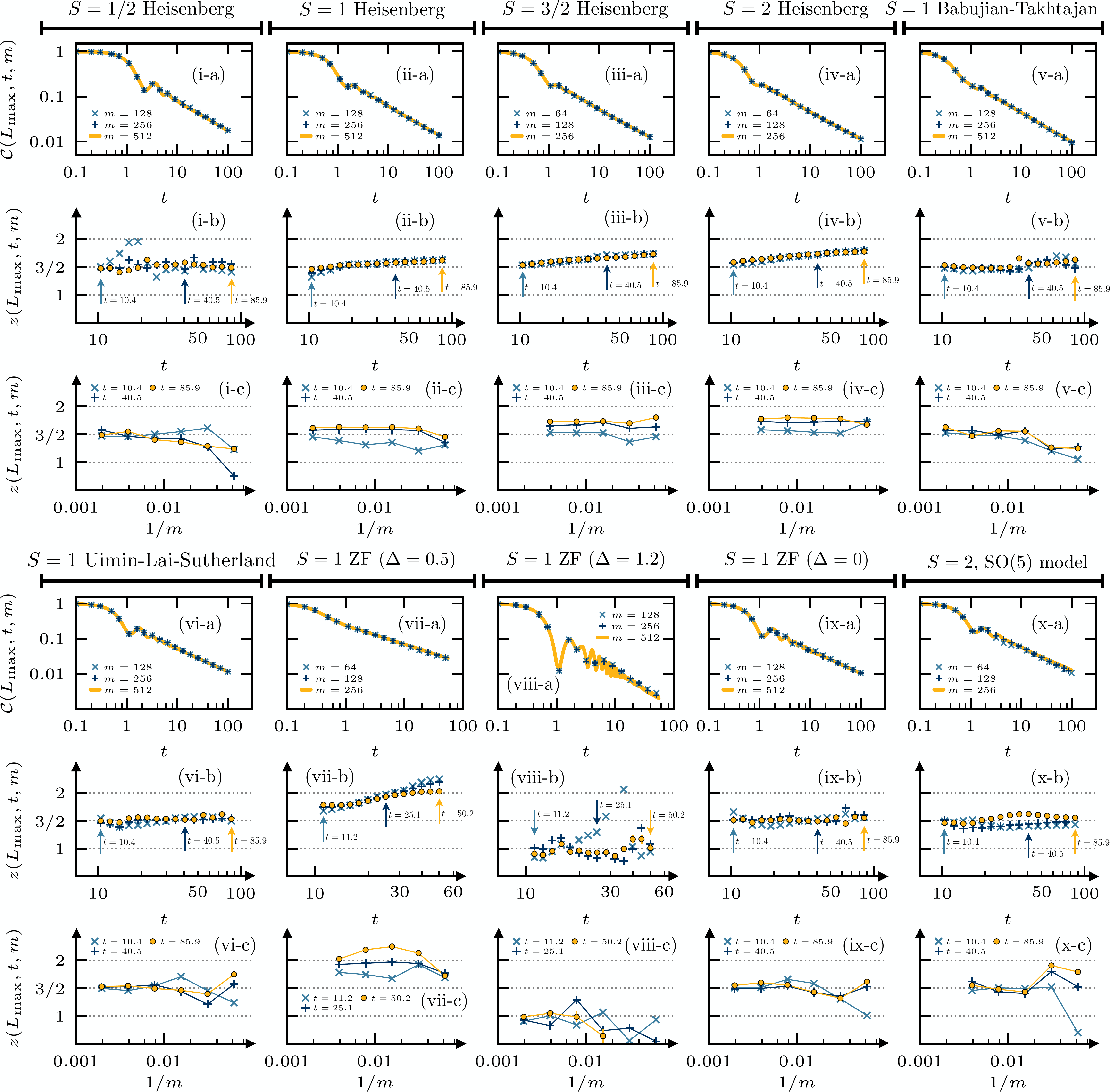}
    \caption{Top rows (a): Infinite temperature spin-spin correlation function for the largest system size $L_\mathrm{max}$ for each of the models considered in the main text (see Tab.~\ref{tab:sim_parameters}) for different bond dimensions $m$. Middle rows (b): For the same three values of the bond dimension of the corresponding upper panel, the dynamical exponent $z$ is extracted by performing curve-fitting inside a sliding window of $10t$ data points, and is plotted versus $t$. Bottom rows (c): For three values of the time $t$ (short, intermediate and long -- and highlighted in the corresponding upper panel by an arrow), the dynamical exponent is fitted inside a window containing $10t$ data points and plotted versus the inverse bond dimension $1/m$. Lines are guide for the eye. The panels correspond to: (i) Heisenberg $S=1/2$, (ii) Heisenberg $S=1$, (iii) Heisenberg $S=3/2$, (iv) Heisenberg $S=2$, (v) Babujian-Takhtajan $S=1$, (vi) Uimin-Lai-Sutherland $S=1$, (vii) Zamolodchikov-Fateev $S=1$ at $\Delta=0.5$, (viii) Zamolodchikov-Fateev $S=1$ at $\Delta=1.2$, (ix) Zamolodchikov-Fateev $S=1$ at $\Delta=0.0$, (x) $\mathrm{SO}(5)$ symmetric $S=2$ model.}
    \label{fig:fit_bond_dimension}
\end{figure*}

For each model considered in the main text, we show in Fig.~\ref{fig:fit_bond_dimension} that for the largest system size $L_\mathrm{max}$ (see Tab.~\ref{tab:sim_parameters}) good convergence versus the bond dimension $m$ is achieved for the extracted dynamical exponent $z$. Respectively in the (b) middle and (c) bottom rows of Fig.~\ref{fig:fit_bond_dimension}, we display $z$ versus time for three values of the bond dimension and $z$ versus the inverse bond dimension for three values of the time (short, intermediate and long).

First excluding the $S=1$, $S=3/2$ and $S=2$ Heisenberg models, we observe in Fig.~\ref{fig:fit_bond_dimension} a systematic convergence of $z$ in the limit $t\to\infty$ and $1/m\to 0$ to either $z=1$, $z=3/2$ or $z=2$, depending on the case. In particular, the dynamical exponent takes one of these three values and not something in between, random or out of control. Plus, our results are consistent with one another depending on the properties of the models (e.g., integrable, non-integrable, isotropic). Based on this, one can then argue that for the $S=1$, $S=3/2$ and $S=2$ Heisenberg models, the numerics should also be reliable (convergence is indeed observed as $1/m\to 0$). Computationally, we are not able to reach long enough times to observe convergence as $t\to\infty$. This means that there is a relatively long crossover before reaching the asymptotic long-time limit, which is going to be diffusive since it looks like $z\to 2$ as $t\to\infty$. As discussed in the main text, such a relatively long crossover also exists for the classical Heisenberg model. Although smaller (hence we are able to resolve it), a crossover is also visible for the non-integrable ZF model at $\Delta=1.2$ in Fig.~\ref{fig:fit_bond_dimension}\;(vii-b) as well as for the non-integrable $S=1$ and $S=3/2$ XY models of Fig.~\ref{fig:extra_models_suppl}\;(1-2).

In fact, it is very interesting that a moderate bond dimension seems sufficient to accurately capture the correct algebraic behavior $\sim t^{-1/z}$ at long time at infinite temperature. It is unclear why entanglement (the amount of entanglement that can be encoded is controlled by the bond dimension $m$, and which is therefore bounded by $m_\mathrm{max}$) has little to do with it, but this surely opens perspectives for future studies (see  also Ref.~\onlinecite{leviatan2017}). We also want to mention that other works, see e.g., Refs.~\onlinecite{ljubotina2017},~\onlinecite{ljubotina2019} and~\onlinecite{varma2019}, successfully addressing similar questions, use a finite bond dimension that is way smaller than one would naively require for the system sizes and times considered.

\section*{Finite-size effects}

When studying long-time dynamics on a finite system, the system size $L$ has to be large compared to the causality light cone to avoid any finite size effect. In order to take this into account, the data for the smaller system sizes are shown until the time at which a significant deviation from the larger system size is visible, while we typically consider $t_\mathrm{max}\sim 100$ otherwise.

For instance, if one considers Fig.~1\;(d,\,f,\,h) of the main text, there is a perfect collapse of $L=32$ data onto the $L=64$ data, which also collapse nicely onto the $L=128$ data. This collapse survives for later and later times as the system size is increased. The deviation that can be observed at ``long time'' for the small system sizes is a causality light cone effect as this is only observed at longer times for larger system sizes. Within the light cone, there is no systematic deviation from small to large system sizes for the exponent.

On Fig.~2\;(b) of the main text, there is at all time a systematic deviation of the data from $L=32$ to $L=128$, with no overlap for the value of the exponent (even within a time window within the causality light cone). But as one considers larger and larger system sizes $L$, the exponent goes toward $z=3/2$, converging to its thermodynamic value.

\section*{Additional models}

\begin{figure*}[!b]
    \centering
    \includegraphics[width=0.9\columnwidth]{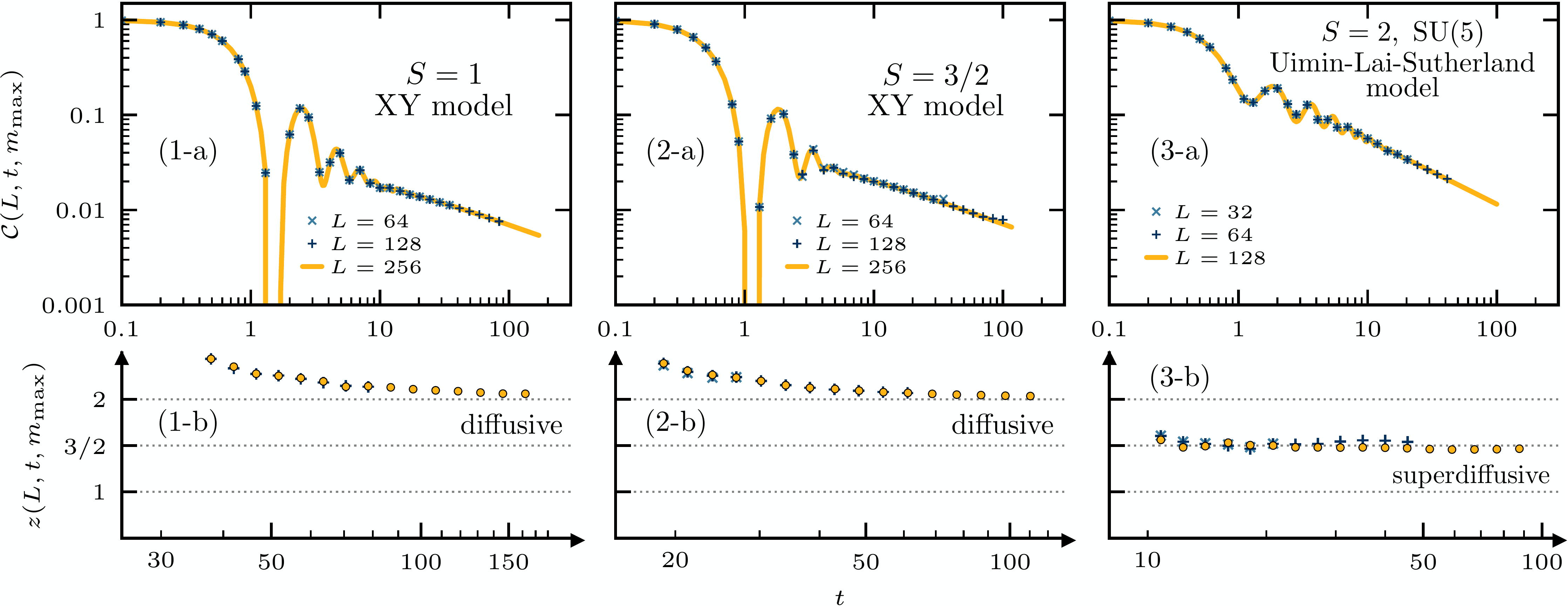}
    \caption{The left column corresponds to the $S=1$ XY model of Eq.~\eqref{eq:xy_s}, the middle one to the $S=3/2$ XY model of Eq.~\eqref{eq:xy_s} and the right one to the $\mathrm{SU}(5)$ Uimin-Lai-Sutherland model of Eq.~(6) of the main text with $\theta=\arctan(1/3)$. Top panels: Infinite temperature spin-spin autocorrelation function. Bottom panels: Extracted dynamical exponent $z(L,t,m_\mathrm{max})$ by performing curve-fitting inside a sliding window of data points in order to reliably the infinite-length and infinite-time value of the power-law decay. For the first two columns, diffusive behavior is observed with $z(L,t,m_\mathrm{max})\to 2$ at long time. In the last column, superdiffusion is obtained with $z(L,t,m_\mathrm{max})=3/2$, as expected for an isotropic and integrable model.}
    \label{fig:extra_models_suppl}
\end{figure*}

\begin{figure*}[!h]
    \centering
    \includegraphics[width=\columnwidth]{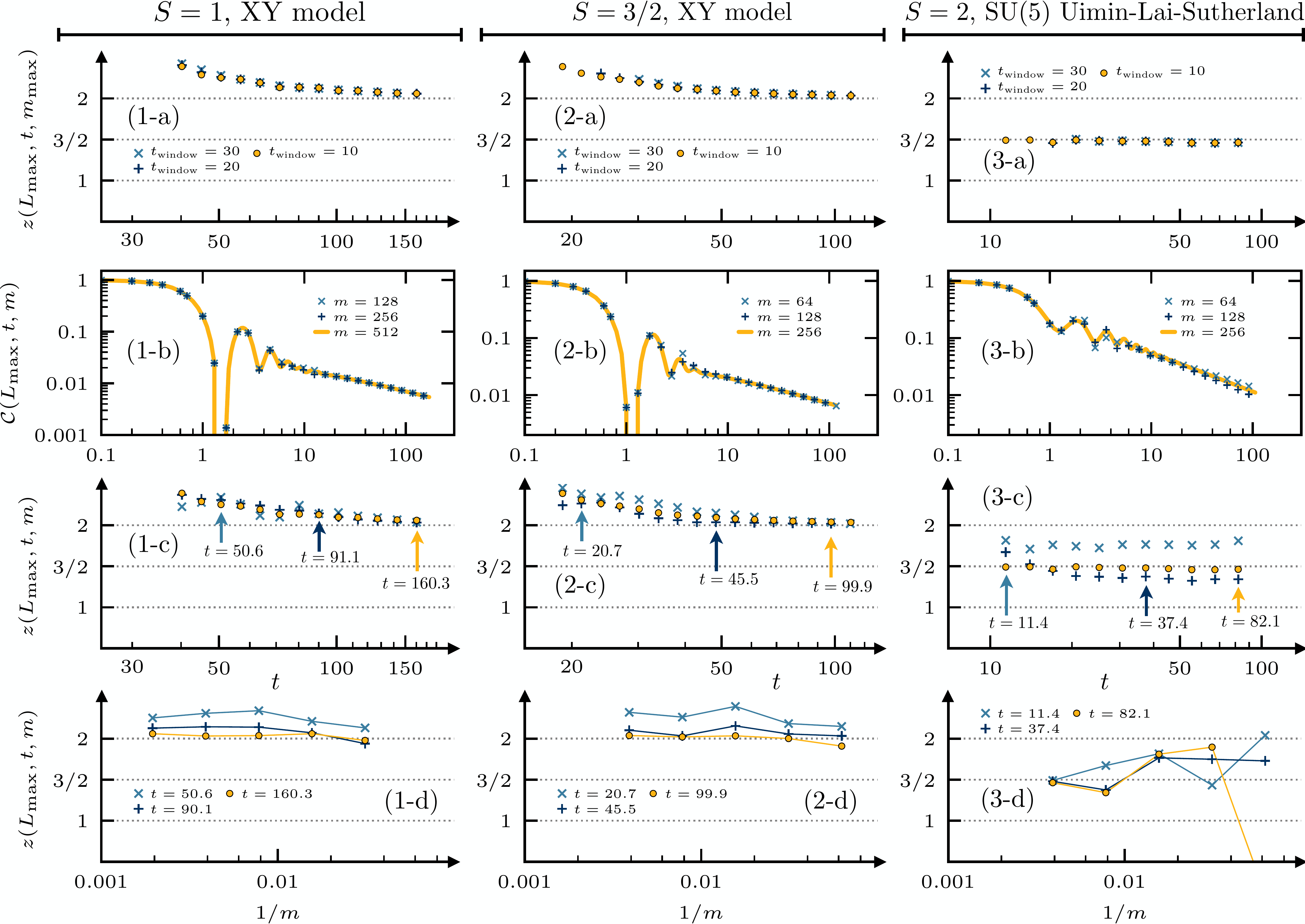}
    \caption{The left column corresponds to the $S=1$ XY model of Eq.~\eqref{eq:xy_s}, the middle one to the $S=3/2$ XY model of Eq.~\eqref{eq:xy_s} and the right one to the $\mathrm{SU}(5)$ Uimin-Lai-Sutherland model of Eq.~(6) of the main text with $\theta=\arctan(1/3)$. Panels (a) display the effect of the size of the fitting window ($10$, $20$ and $30$) on the extracted dynamical exponent. Panels (b) show the correlation function for the largest system size considered (see Tab.~\ref{tab:sim_parameters}) for various bond dimensions. Panels (c) show the extracted dynamical exponent from the data of panels (d) by performing curve-fitting inside a sliding window. Panels (e) display the value of the extracted dynamical exponent versus the inverse bond dimension for the largest system size considered, at three different times (short, intermediate and long).}
    \label{fig:extra_models_conv_suppl}
\end{figure*}

In addition to the ten models considered in the main text, we consider four extra models in this supplemental material, strengthening our conclusions. First, we look at the XY $S=1$ and $S=3/2$ models, which are non-integrable and described by the local Hamiltonian density,
\begin{align}
    \hat{h}_{j,j+1}=S^x_jS^x_{j+1} + S^y_jS^y_{j+1}.
    \label{eq:xy_s}
\end{align}
Then, we study the integrable $\mathrm{SU}(5)$ Uimin-Lai-Sutherland model through a spin-$2$ representation. Its Hamiltonian is the same as Eq.~(6) of the main text for the value of the parameter $\theta=\arctan(1/3)$. The infinite-temperature local spin-spin correlation function is computed similarly to all other models, and the data are displayed in Fig.~\ref{fig:extra_models_suppl}. As expected, we get superdiffusive spin dynamics for the $\mathrm{SU}(5)$ integrable model. It is diffusive for the other cases, and there exists a finite crossover time before reaching the asymptotic long-time limit. The dynamical exponent at short time [Fig.~\ref{fig:extra_models_suppl}\;(1-a) and (2-a)] takes a value $z>2$ before reaching $z\to 2$ at long time, despite the integrable low-energy effective theory describing the Hamiltonian~\eqref{eq:xy_s}. No trace of ballistic behavior is observed in the dynamical exponent $z(t)$, while this is what one would expect if the low-energy theory played a role in the long-time dynamics at infinite-temperature. This is an interesting observation in regards of the isotropic and non-integrable Heisenberg model with $S\geq 1$ studied in the main text (which also has an integrable low-energy effective theory). Indeed, it displayed a short-time exponent $z\approx 3/2$, which could be misleading to distinct diffusive from superdiffusive dynamics. Additional analyses to reliably extract the dynamical exponent are available in Fig.~\ref{fig:extra_models_conv_suppl}.

\begin{figure*}[!h]
    \centering
    \includegraphics[width=0.9\columnwidth]{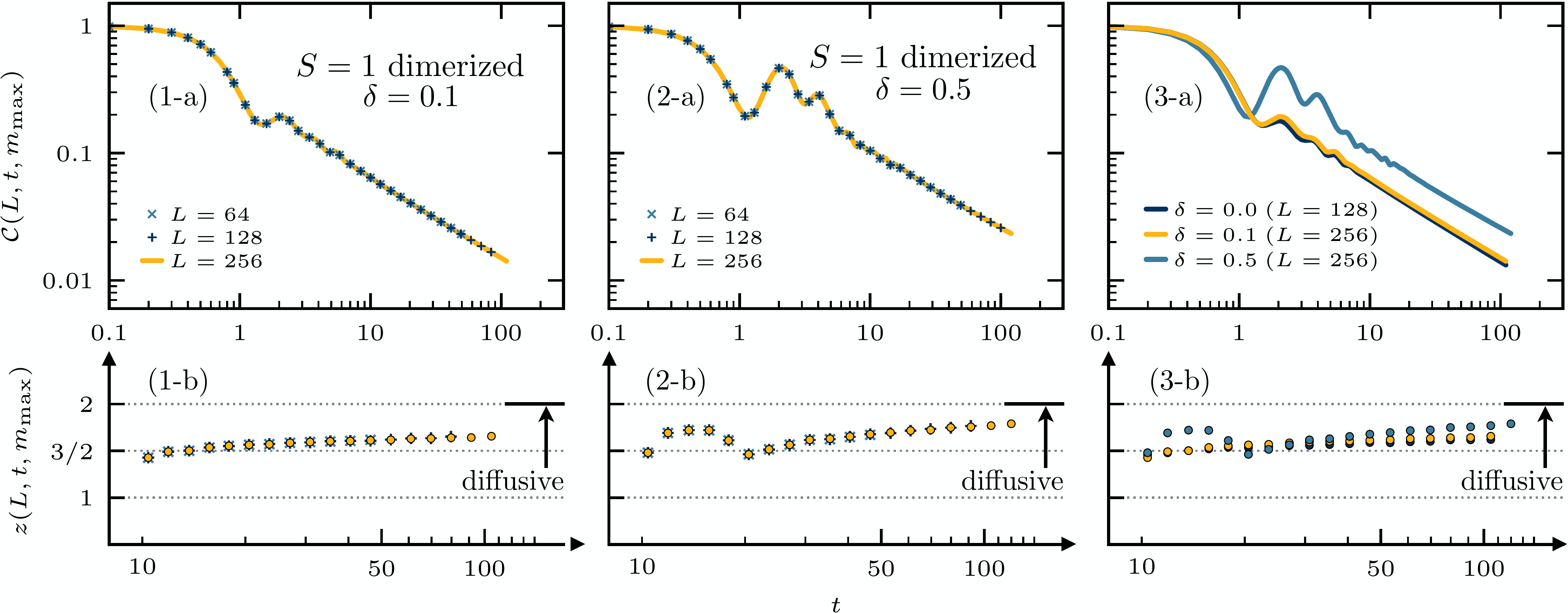}
    \caption{The two left columns correspond to the dimerized spin-$1$ chain of Eq.~\eqref{eq:s1_dimerized} at $\delta=0.1$ and $\delta=0.5$ respectively. The right panel is a direct comparison for $\delta=0$, $\delta=0.1$ and $\delta=0.5$. Top panels: Infinite temperature spin-spin autocorrelation function. Bottom panels: Extracted dynamical exponent $z(L,t,m_\mathrm{max})$ by performing curve-fitting inside a sliding window of data points in order to reliably the infinite-length and infinite-time value of the power-law decay. Diffusive behavior is expected with $z(L,t,m_\mathrm{max})\to 2$ as $t\to\infty$ but the long crossover times prevent its definite observation. However, as one increases the value of $\delta$, the crossover time is reduced.}
    \label{fig:s1_af_dimerized_suppl}
\end{figure*}

\begin{figure*}[!h]
    \centering
    \includegraphics[width=\columnwidth]{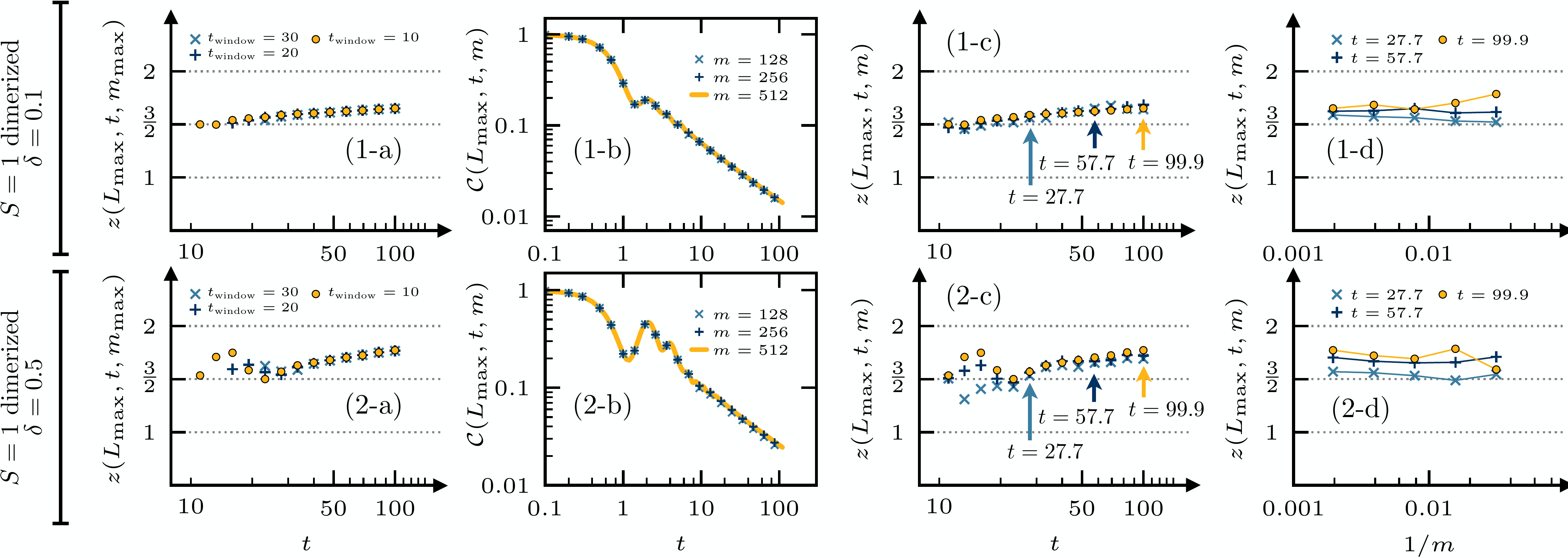}
    \caption{The first row corresponds to the isotropic dimerized $S=1$ model of Eq.~\eqref{eq:s1_dimerized} at $\delta=0.1$, and the second row to the same model at $\delta=0.5$. Panels (a) display the effect of the size of the fitting window ($10$, $20$ and $30$) on the extracted dynamical exponent. Panels (b) show the correlation function for the largest system size considered (see Tab.~\ref{tab:sim_parameters}) for various bond dimensions. Panels (c) show the extracted dynamical exponent from the data of panels (d) by performing curve-fitting inside a sliding window. Panels (e) display the value of the extracted dynamical exponent versus the inverse bond dimension for the largest system size considered, at three different times (short, intermediate and long).}
    \label{fig:s1_af_dimerized_conv_suppl}
\end{figure*}

The last model considered is the non-integrable isotropic dimerized $S=1$ chain described by the local Hamiltonian density,
\begin{align}
    \hat{h}_{j,j+1}=\Bigl[1 + \left(-1\right)^j\delta\Bigr]\boldsymbol{S}_j\cdot\boldsymbol{S}_{j+1},
    \label{eq:s1_dimerized}
\end{align}
with $\delta>0$ controlling the strength of the dimerization between even and odd bonds. Its ground state belongs to the Haldane phase for $\delta\lesssim 0.25$ and to a dimerized phase for larger values~\cite{kitazawa1997}. The infinite-temperature local spin-spin correlation function is displayed in Fig.~\ref{fig:s1_af_dimerized_suppl} for $\delta=0.1$ and $\delta=0.5$, with diffusive behavior expected at long time. As for the $S=1$ Heisenberg chain considered in the main text ($\delta=0$), there seems to be relatively long crossover before reaching the asymptotic diffusive dynamics. However, as one increases the value of $\delta$, this crossover time is reduced, as shown in Fig.~\ref{fig:s1_af_dimerized_suppl}\;(3). Since both the Haldane and dimerized phases can be described by the non-linear sigma model in the low-energy limit, this confirms one more that the (integrable) low-energy effective field theory plays no role in the dynamics in the long time limit at infinite temperature. Additional analyses to reliably extract the dynamical exponent are available in Fig.~\ref{fig:s1_af_dimerized_conv_suppl}.

\newpage
\section*{Summary of the parameters}

\begin{table}[!h]
    \begin{minipage}{0.8\columnwidth}
        \center
        \begin{ruledtabular}
            \begin{tabular}{cccc}
                \thead{Model} & \thead{System size $L$} & \thead{Maximum bond dimension considered $m_\mathrm{max}$}\\
                \hline
                \multirow{4}{*}{\makecell{\thead{Heisenberg\\ $S=1/2$}}}
                & \makecell{$32$} & \makecell{$2048$}\\
                & \makecell{$64$} & \makecell{$2048$}\\
                & \makecell{$128~(L_\mathrm{max})$} & \makecell{$512$}\\
                \hline
                \multirow{4}{*}{\makecell{\thead{Heisenberg\\ $S=1$}}}
                & \makecell{$32$} & \makecell{$1024$}\\
                & \makecell{$64$} & \makecell{$1024$}\\
                & \makecell{$128~(L_\mathrm{max})$} & \makecell{$512$}\\
                \hline
                \multirow{4}{*}{\makecell{\thead{Heisenberg\\ $S=3/2$}}}
                & \makecell{$32$} & \makecell{$512$}\\
                & \makecell{$64$} & \makecell{$512$}\\
                & \makecell{$128~(L_\mathrm{max})$} & \makecell{$256$}\\
                \hline
                \multirow{4}{*}{\makecell{\thead{Heisenberg\\ $S=2$}}}
                & \makecell{$32$} & \makecell{$512$}\\
                & \makecell{$64$} & \makecell{$512$}\\
                & \makecell{$128~(L_\mathrm{max})$} & \makecell{$256$}\\
                \hline
                \multirow{4}{*}{\makecell{\thead{Babujian-Takhtajan\\ $S=1$}}}
                & \makecell{$32$} & \makecell{$1024$}\\
                & \makecell{$64$} & \makecell{$1024$}\\
                & \makecell{$128~(L_\mathrm{max})$} & \makecell{$512$}\\
                \hline
                \multirow{4}{*}{\makecell{\thead{Uimin-Lai-Sutherland\\ $S=1$}}}
                & \makecell{$32$} & \makecell{$1024$}\\
                & \makecell{$64$} & \makecell{$1024$}\\
                & \makecell{$128~(L_\mathrm{max})$} & \makecell{$512$}\\
                \hline
                \multirow{4}{*}{\makecell{\thead{Zamolodchikov-Fateev\\ $S=1$, $\Delta=0.5$}}}
                & \makecell{$32$} & \makecell{$1024$}\\
                & \makecell{$64$} & \makecell{$1024$}\\
                & \makecell{$128~(L_\mathrm{max})$} & \makecell{$512$}\\
                \hline
                \multirow{4}{*}{\makecell{\thead{Zamolodchikov-Fateev\\ $S=1$, $\Delta=1.2$}}}
                & \makecell{$128$} & \makecell{$512$}\\
                & \makecell{$256$} & \makecell{$256$}\\
                & \makecell{$512~(L_\mathrm{max})$} & \makecell{$256$}\\
                \hline
                \multirow{4}{*}{\makecell{\thead{Zamolodchikov-Fateev\\ $S=1$, $\Delta=0$}}}
                & \makecell{$32$} & \makecell{$1024$}\\
                & \makecell{$64$} & \makecell{$1024$}\\
                & \makecell{$128~(L_\mathrm{max})$} & \makecell{$512$}\\
                \hline
                \multirow{4}{*}{\makecell{\thead{SO(5)-symmetric\\ $S=2$, $\theta=\mathrm{arctan}(1/9)$}}}
                & \makecell{$32$} & \makecell{$512$}\\
                & \makecell{$64$} & \makecell{$512$}\\
                & \makecell{$128~(L_\mathrm{max})$} & \makecell{$256$}\\
                \hline
                \multirow{4}{*}{\makecell{\thead{SU(5)-symmetric\\ $S=2$, $\theta=\mathrm{arctan}(1/3)$}}}
                & \makecell{$32$} & \makecell{$512$}\\
                & \makecell{$64$} & \makecell{$512$}\\
                & \makecell{$128~(L_\mathrm{max})$} & \makecell{$256$}\\
                \hline
                \multirow{4}{*}{\makecell{\thead{XY $S=1$}}}
                & \makecell{$64$} & \makecell{$512$}\\
                & \makecell{$128$} & \makecell{$512$}\\
                & \makecell{$256~(L_\mathrm{max})$} & \makecell{$512$}\\
                \hline
                \multirow{4}{*}{\makecell{\thead{XY $S=3/2$}}}
                & \makecell{$64$} & \makecell{$512$}\\
                & \makecell{$128$} & \makecell{$256$}\\
                & \makecell{$256~(L_\mathrm{max})$} & \makecell{$256$}\\
        \end{tabular}
    \end{ruledtabular}
\end{minipage}
\end{table}
\begin{table}[!h]
    \begin{minipage}{0.8\columnwidth}
        \center
        \begin{ruledtabular}
            \begin{tabular}{cccc}
                \multirow{4}{*}{\makecell{\thead{Dimerized $S=1$\\$\delta=0.1$}}}
                & \makecell{$64$} & \makecell{$512$}\\
                & \makecell{$128$} & \makecell{$512$}\\
                & \makecell{$256~(L_\mathrm{max})$} & \makecell{$512$}\\
                \hline
                \multirow{4}{*}{\makecell{\thead{Dimerized $S=1$\\$\delta=0.5$}}}
                & \makecell{$64$} & \makecell{$512$}\\
                & \makecell{$128$} & \makecell{$512$}\\
                & \makecell{$256~(L_\mathrm{max})$} & \makecell{$512$}\\
            \end{tabular}
        \end{ruledtabular}
        \caption{Maximum bond dimension $m_\mathrm{max}$ used in the simulation of the different systems of length $L$ considered in the main text and in the supplemental material. For each system, the maximum size $L_\mathrm{max}$ considered is also highlighted.}
        \label{tab:sim_parameters}
    \end{minipage}
\end{table}

\end{document}